\documentclass[english,english,aps, prc, manuscript,superscriptaddress,reprint,showpacs,amsmafixth,amssymb,float]{revtex4-1}
\usepackage{mathptmx}

\usepackage[T1]{fontenc}
\usepackage[latin9]{inputenc}
\usepackage{bm}
\usepackage{graphicx}
\usepackage{esint}

\makeatletter

\providecommand{\tabularnewline}{\\}

 \@ifundefined{textcolor}{}
 {%
   \definecolor{BLACK}{gray}{0}
   \definecolor{WHITE}{gray}{1}
   \definecolor{RED}{rgb}{1,0,0}
   \definecolor{GREEN}{rgb}{0,1,0}
   \definecolor{BLUE}{rgb}{0,0,1}
   \definecolor{CYAN}{cmyk}{1,0,0,0}
   \definecolor{MAGENTA}{cmyk}{0,1,0,0}
   \definecolor{YELLOW}{cmyk}{0,0,1,0}
 }

\usepackage{mathptmx}

\usepackage[T1]{fontenc}
\usepackage[latin9]{inputenc}
\usepackage{geometry}
\usepackage{array}
\usepackage{bm}
\usepackage{graphicx}
\usepackage{esint}
\usepackage[dvipdfm,bookmarks=true,colorlinks,%
            citecolor=blue,linkcolor=blue,hypertex, %
            breaklinks=true]{hyperref}
\makeatletter

\providecommand{\tabularnewline}{\\}

 \@ifundefined{textcolor}{}
 {%
   \definecolor{BLACK}{gray}{0}
   \definecolor{WHITE}{gray}{1}
   \definecolor{RED}{rgb}{1,0,0}
   \definecolor{GREEN}{rgb}{0,1,0}
   \definecolor{BLUE}{rgb}{0,0,1}
   \definecolor{CYAN}{cmyk}{1,0,0,0}
   \definecolor{MAGENTA}{cmyk}{0,1,0,0}
   \definecolor{YELLOW}{cmyk}{0,0,1,0}
 }

\usepackage{color}
\usepackage{graphicx}
\usepackage{bm}
\usepackage{times}

\makeatother

\usepackage{babel}

\makeatother

\usepackage{babel}
\begin{document}

\title{Breaking and restoration of rotational symmetry for irreducible tensor
operators on the lattice}

\author{Bing-Nan Lu}

\address{Institute for Advanced Simulation, Institut für Kernphysik, and Jülich
Center for Hadron Physics, Forschungszentrum Jülich, D-52425 Jülich,
Germany}

\author{Timo A. Lähde}

\address{Institute for Advanced Simulation, Institut für Kernphysik, and Jülich
Center for Hadron Physics, Forschungszentrum Jülich, D-52425 Jülich,
Germany}

\author{Dean Lee}

\address{Department of Physics, North Carolina State University, Raleigh,
North Carolina 27695, USA}

\author{Ulf-G. Meißner}

\address{Institute for Advanced Simulation, Institut für Kernphysik, and Jülich
Center for Hadron Physics, Forschungszentrum Jülich, D-52425 Jülich,
Germany}

\address{Helmoholtz-Institut für Strahlen- und Kernphysik and Bethe Center
for Theoretical Physics, \\
Universität Bonn, D-53115 Bonn, Germany}

\address{JARA-High Performance Computing, Forschungszentrum Jülich, D-52425
Jülich, Germany}

\date{4-March-2015}
\begin{abstract}
We study the breaking of rotational symmetry on the lattice for
irreducible tensor operators and practical methods for suppressing
this breaking. We illustrate the features of the general problem using an $\alpha$
cluster model for $^{8}$Be. We focus on the lowest states with non-zero
angular momentum and examine the matrix elements of multipole moment operators.
We show that the physical reduced matrix
element is well reproduced by averaging over all possible orientations of the quantum state, and this is expressed as a sum of matrix elements weighted by the corresponding Clebsch-Gordan coefficients. For our $\alpha$
cluster model we find that the effects of rotational symmetry breaking
can be largely eliminated for lattice spacings of $a\leq 1.7$ fm, and we expect similar improvement for actual lattice Monte Carlo calculations.
\end{abstract}

\pacs{12.38.Gc, 03.65.Ge, 21.10.Dr}

\maketitle

\section{introduction}

In recent years, lattice Monte Carlo calculations have been widely
applied to the study of nuclear structure~\cite{Lee2009_PPNP63-117,Bazavov2010_RMP82-1349,Beane2011_PPNP66-1}.
In particular, chiral effective field theory (chiral EFT) combined with
lattice methods has been employed to study the spectrum and structure
of light and medium-mass nuclei~\cite{Borasoy2007_EPJA31-105,Epelbaum2010_EPJA45-335,Epelbaum2010_PRL104-142501,Epelbaum2011_PRL106-192501,Epelbaum2013_PRL110-112502}.
In such calculations, continuous space-time is discretized and compactified
so that path integrals can be computed numerically. The mesh points
uniformly span a cubic box, and some boundary conditions such as periodic
boundaries are imposed in each dimension. However, the calculated bound
state energies and wave functions will, in general, deviate from their
continuum infinite-volume counterparts due to the discretization and
finite-volume artifacts. 

Over the years, much effort has
been devoted to removing numerical artifacts in lattice field theory calculations. The finite volume energy shifts for two-body bound states~\cite{Luescher1986_CMP104-177,Luescher1991_NPB354-531,Beane2004_PLB585-106,Koenig2011_PRL107-112001,Koenig2012_AP327-1450,Bour2011_PRD84-091503,Davoudi2011_PRD84-114502,Briceno2013_arXiv1311-7686,Briceno2013_PRD88-114507,Agadjanov2014_JHEP2014-1}
as well as two-body resonances and scattering problems~\cite{Bernard2008_JHEP2008-24,Lage2009_PLB681-439,Bernard2011_JHEP2011-1,Gockeler2012_PRD86-094513,Li2013_PRD87-014502} have been studied in detail. There is also on-going research to extend  these results to bound states with more
than two constituents~\cite{Koenig2011_PRL107-112001,Koenig2012_AP327-1450,Kreuzer2010_EPJA43-229,Kreuzer2011_PLB694-424,Polejaeva2012_EPJA48-1,Briceno2013_PRD87-094507,Meissner2015_PRL114-091602}.
On the other hand, removing the artifacts from finite lattice spacing
is a more complicated issue. For chiral EFT the lattice improvement
program proposed by Symanzik\textit{ et al}.~\cite{Weisz1983_NPB212-1,Weisz1984_NPB236-397,Luescher1984_NPB240-349}
provides a useful approach for systematically reducing discretization
errors. This method was also applied to Yang-Mills theories~\cite{Weisz1983_NPB212-1,Weisz1984_NPB236-397},
gauge field theories~\cite{Luescher1984_NPB240-349,Curci1983_PLB130-205,Hamber1983_PLB133-351,Eguchi1984_NPB237-609}
and QCD~\cite{Sheikholeslami1985_NPB259-572}. Dudek \textit{et al}.~\cite{Dudek2009_PRL103-262001}
have proposed a method where the continuum spin of meson~\cite{Dudek2010_PRD82-034508}
and baryon~\cite{Edwards2011_PRD84-074508,Meinel2012_PRD85-114510}
excited states in lattice QCD can be reliably identified. Meanwhile,
Davoudi \textit{et al}.~\cite{Davoudi2012_PRD86-054505} have quantified
the breaking and restoration of rotational invariance at both tree
level and one-loop level by means of lattice operators smeared over
a finite spatial region.

On the lattice, the full rotational symmetry group is reduced to the
finite group of cubic rotations, $SO(3) \rightarrow SO(3,Z)$. Several basic rules based on the
argument of rotational invariance are broken. For example, in the continuum
and infinite-volume limits, quantum bound states with angular momentum
$J$ form a degenerate multiplet consisting of $2J+1$ components, while
on the lattice the energy levels split into subgroups corresponding
to different irreducible representations (\textit{irreps}) of the
cubic group~\cite{Johnson1982_PLB114-147,Berg1983_NPB221-109,Mandula1983_NPB228-91}.
The size of the energy splittings are dictated by the lattice spacing
and by the volume and boundary conditions.

In Ref.~\cite{Lu2014_PRD90-034507}, we explored the breaking of
rotational symmetry on the lattice for bound state energies with an
$\alpha$ cluster model. It was shown that the calculated energy is
minimized when the natural separation between particles is commensurate
with the separation between lattice points along the preferred lattice
directions associated with the given angular momentum state. It was also shown that the multiplet-averaged energy is
closer to the continuum limit than any single energy level. One can apply these results to future \textit{ab
initio} lattice simulations for nuclear systems where $\alpha$ cluster
structures are important.

In this paper, we extend the analysis of Ref.~\cite{Lu2014_PRD90-034507}
to other observables besides the energy. One is often interested in the nuclear radii, quadrupole moments as well
as transition probabilities. For example, an anomalously large 
radius compared with the usual $A^{1/3}$ scaling law is evidence for a halo
nucleus~\cite{Tanihata1985_PRL55-2676,Meng2006_PPNP57-470}, while
the intrinsic quadrupole moment is often related to the rotational
bands observed in deformed nuclei~\cite{Bohr1976_RMP48-365}.

We consider irreducible tensor operators sandwiched by a pair of
bound state wave functions. In the continuum limit such an expression
can be factorized and simplified according to the Wigner-Eckart theorem
because of the full rotational symmetry, but on the lattice such
factorization is no longer possible and the situation becomes more complicated. Furthermore, continuum selection rules for electromagnetic
transitions are not exactly satisfied on the lattice. Some transitions that are absolutely forbidden by rotational symmetry 
may assume non-zero amplitudes on the lattice. We would like to construct proper corrections
for these matrix elements in order to minimize symmetry breaking
effects.

Our objective in this paper is to investigate anisotropic lattice artifacts in the matrix elements of irreducible tensor
operators and search for a practical method to restore rotational
symmetry. The details of the particular interaction are not essential
to our general analysis. Here, we use the same $\alpha$ cluster model as
in Ref.~\cite{Lu2014_PRD90-034507} where the $\alpha$-$\alpha$
interaction is approximated by an Ali-Bodmer type potential adjusted to produce a bound  $^{8}$Be nucleus.

\section{\label{sec:II}theoretical framework}

\subsection{Hamiltonian}

Let $m_{\alpha}=3727.0$ MeV denote the mass of the $\alpha$ particle
and $m=m_{\alpha}/2$ the reduced mass. Our starting point is the
one-body Hamiltonian 
\begin{equation}
H=-\frac{\nabla^{2}}{2m}+V(r),\label{eq:continuumHamiltonian}
\end{equation}
where $r=\sqrt{x^{2}+y^{2}+z^{2}}$ is the distance between the two
$\alpha$ particles, $V=V_{N}+V_{C}$ is the $\alpha$-$\alpha$ potential,
including nuclear and Coulomb potentials. 

We use the same two-body potential as that used in Ref.~\cite{Lu2014_PRD90-034507}.
For completeness we briefly introduce its functional form and the
parameters. For  the nuclear part of the $\alpha$-$\alpha$ interaction
we use an isotropic Ali-Bodmer-type potential,

\begin{equation}
	V_{N}(r)=V_{0}\exp{(-\eta_{0}^{2}r^{2})}+V_{1}\exp{(-\eta_{1}^{2}r^{2})},
\end{equation}
where $V_{0}=-216.0$ MeV, $V_{1}=354.0$ MeV, $\eta_{0}=0.436$
fm$^{-1}$ and $\eta_{1}=0.529$ fm$^{-1}$. These parameters are
determined by fixing the $S$- and $D$-wave $\alpha$-$\alpha$
scattering lengths to their experimental values. The Coulomb potential
is given by the error function
\begin{equation}
V_{C}(r)=\frac{4e^{2}}{r}{\rm erf}\left(\frac{\sqrt{3}r}{2R_{\alpha}}\right),
\end{equation}
where $R_{\alpha}=1.44$ fm is the radius of the $\alpha$ particle,
$e$ is the fundamental unit of charge and erf denotes the error function. 

Since our focus is on measuring bound state properties and the physical $^{8}$Be nucleus is unbound, we increase $V_{0}$ by
an amount of 30\%.
With this strengthened potential, the $^{8}$Be nucleus has a ground 
state at $E(0^{+})=-10.8$~MeV and one excited state at $E(2^{+})=-3.3$~MeV. 
These energies are measured relative to the $\alpha$-$\alpha$
threshold. 

In the lattice calculations the spatial vector $\bm{r}$ assumes discrete
values, and so the Hamiltonian in Eq. (1) becomes a matrix. Periodic boundary conditions
for a box of size $L$ are imposed on the wave functions,
\begin{equation}
\psi(\bm{r}+\bm{n}_{i}L)=\psi(\bm{r}),
\end{equation}
where $\bm{n}_{i}$ with $i=x,y,z$ are the unit vectors along the
three coordinate axes. The energy eigenvalues and wave functions can be
obtained by the diagonalization of the Hamiltonian matrix of dimension
$L^{3(N-1)}\times L^{3(N-1)}$.

The kinetic energy term in Eq. (1) can be expressed on the lattice
by finite differences. In one dimension we have
\begin{equation}
f^{\prime\prime}(x)\approx c_{0}^{(N)}f(x)+\sum_{k=1}^{N}c_{k}^{(N)}\left[f(x+ka)+f(x-ka)\right],
\end{equation}
where $a$ is the lattice spacing and $c_{k}^{(N)}$ is a set of fixed
coefficients. The order-$N$ formula involves $2N+1$ points and the
corresponding truncation error is $\mathcal{O}(a^{2N})$. In Ref.~\cite{Lu2014_PRD90-034507},
we gave the coefficients $c_{k}^{(N)}$ for $N\leq4$. In this paper,
we use the $N=4$ formula to calculate the second-order derivatives
appearing in the Laplace operator. This choice removes most of the
rotational symmetry breaking effects due to the difference formula for the kinetic energy.

\subsection{\label{sub:2-b}Lattice wave functions}

The continuum Hamiltonian Eq.~(\ref{eq:continuumHamiltonian}) is
invariant under any spatial rotation. As a result, the bound states
of $H$ form angular momentum multiplets.
Let us denote the bound state wave
functions as $\phi_{lm}$, where the integer $l$ is the angular momentum and  the integer $m$ is the $z$-component of angular momentum with $-l\leq m\leq l$. For systems with more than one bound state
with the same  value of $l$, we will need additional
radial quantum numbers. However, we do not consider such  cases here. The angular dependence of these wave functions
are given by the corresponding spherical harmonics $Y_{lm}$.

On the lattice, these angular momentum $l$ multiplets are split into  \textit{irreps} of the cubic rotational group, $SO(3,Z)$. In Ref.~\cite{Lu2014_PRD90-034507},
we gave the splitting patterns of the angular momentum multiplets
with $l\leq8$. In order to specify the wave functions belonging to
the same \textit{irrep}, we define a quantum number $k$ that is valid
on the lattice through the relation
\begin{equation}
R_{z}\left(\frac{\pi}{2}\right)=\exp\left(-i\frac{\pi}{2}k\right),
\end{equation}
where $R_{z}(\pi/2)$ is a rotation around the $z$-axis by $\pi/2$.
The integers $k$ are equal to $m$ modulo 4 and are non-degenerate for each\textit{ irrep} of $O$. We label
the wave function $\psi_{l\tau k}$ for any eigenstate by $l$, $k$ and the\textit{ irrep} $\tau$ it belongs to.
If the angular momentum $J$ contains more than one branch belonging
to the same \textit{irrep}, we distinguish them by adding primes to
the name of the \textit{irreps}. For instance, the symbol ``$\psi_{6T_{2}^{\prime}1}$''
denotes the wave function with $l=6$, $k=1$ belonging to the second
$T_{2}$ \textit{irrep}. 
Consequently, in the continuum limit the wave functions $\psi_{l\tau k}$
form a complete basis for the bound state subspace, and the corresponding energies are degenerate for the same total
angular momentum $J$, while on the lattice the energy eigenvalue depends
on both $J$ and the \textit{irrep} $\tau$.
In the continuum limit we can write down unitary transformations from the $\phi_{lm}$
basis to the $\psi_{l\tau k}$ basis and vice versa,
\begin{eqnarray}
\phi_{lm} = \sum_{\tau k} U_{lm\tau k}\psi_{l\tau k}, \label{phi} \\
\psi_{l\tau k} = \sum_{m} U^{-1}_{l\tau km}\phi_{lm} \label{psi}.
\end{eqnarray}
See Ref.~\cite{Altmann:1957}
for details of these transformations. As an example, we show the case of $l=2$. The wave functions $\psi_{2E0}$,
$\psi_{2E2}$ belong to \textit{irrep} $E$, and $\psi_{2T_{2}1}$,
$\psi_{2T_{2}2}$, $\psi_{2T_{2}3}$ belong to \textit{irrep} $T_{2}$.  Following Ref.~\cite{Altmann:1957} we find
\begin{eqnarray}
\psi_{2E0}=\phi_{20},\qquad\psi_{2T_{2}1} & = & \phi_{21},\qquad\psi_{2T_{2}3}=\phi_{2\bar{1}}, \nonumber
\\
\psi_{2E2}=\sqrt{\frac{1}{2}}(\phi_{22}+\phi_{2\bar{2}}) &  & \psi_{2T_{2}2}=-i\sqrt{\frac{1}{2}}(\phi_{22}-\phi_{2\bar{2}}).\label{eq:Jeq2transformation}
\end{eqnarray}
We write $\bar{m}$ for notational convenience to denote $-m$.
  
On the lattice we can obtain the bound state wave functions $\psi_{l\tau k}$ by simultaneously diagonalizing
the lattice Hamiltonian $H$ (or transfer matrix~\cite{Lee2009_PPNP63-117}) and
the $R_{z}(\pi/2)$ operator. Since the full rotational symmetry is broken, the angular
momentum $l$ should be viewed as a label that
describes the angular momentum multiplet obtained by dialing the lattice
spacing continuously to zero.
But we can use the unitary transformation in  Eq.~(\ref{phi})
to define the wavefunctions $\phi_{lm}$ at non-zero lattice spacing.  We do this even though the wavefunctions
$\phi_{lm}$ are generally not exact eigenstates of $H$ when the lattice
spacing is non-zero.  

Consider now the bound state wavefunctions for a zero angular momentum state $\phi_{00}$ and a general state $\phi_{lm}$.  In the continuum limit the matrix elements of a $r^l Y_{lm}$ multipole operator inserted between $\phi_{00}$ and $\phi_{lm}$ must be independent of $m$,
\begin{eqnarray}
\langle\phi_{lm}|r^{l}Y_{lm}|\phi_{00}\rangle \nonumber \\
\equiv\int d^{3}\bm{r} &\langle\phi_{lm}|\rho(\bm{r})r^{l}Y_{lm}(\Omega)|\phi_{00}\rangle=C_{m} \rightarrow C.
\end{eqnarray}
When using Eq.~(\ref{phi}), we find the condition that $C_{m}$ is independent of $m$ a very convenient check that the phases for $\phi_{lm}$ and $\psi_{l\tau k}$ are consistent with standard conventions as defined in Ref.~\cite{Altmann:1957}.

\subsection{\label{sub:2-C}Factorization of the matrix elements}

Given a pair of bound state wave functions $\phi_{l_{1}m_{1}}(\bm{r})$
and $\phi_{l_{2}m_{2}}(\bm{r})$ with angular momenta $l_{1}$, $m_{1}$,
$l_{2}$ and $m_{2}$ on the lattice, we can evaluate the matrix element
of the multipole moment operator $r^{l'}Y_{lm}$,
\begin{eqnarray}
(l_1m_1|r^{l'}Y_{lm}|l_2m_2) & = & \sum_{\bm{n}}\phi_{l_{1}m_{1}}^{*}(\bm{n}a)|\bm{n}a|^{l'^{}}Y_{lm}(\hat{\bm{n}})\phi_{l_{2}m_{2}}(\bm{n}a),\label{eq:matelonlat}
\end{eqnarray}
where $\bm{n}$ runs over all lattice sites. Here, we choose independent
integers $l$ and $l^{\prime}$ in the multipole moment operator,
in order to keep the radial and angular degrees of freedom independent.
This makes our conclusions sufficiently general and applicable to
all irreducible tensor operators. We use parentheses to denote matrix elements on the lattice, $(f|O|i)$, and Dirac brackets to denote matrix elements in the continuum limit, $\langle f|O|i \rangle$.
The continuum limit of Eq.~(\ref{eq:matelonlat}) is 
\begin{equation}
\langle l_{1}m_{1}|r^{l^{\prime}}Y_{lm}|l_{2}m_{2}\rangle=\int d^{3}\bm{r}\phi_{l_{1}m_{1}}^{*}(\bm{r})r^{l^{\prime}}Y_{lm}(\Omega)\phi_{l_{2}m_{2}}(\bm{r})\label{eq:matelincon}
\end{equation}
where the integration is performed over all space. Matrix elements
in the form of Eq.~(\ref{eq:matelincon}) occur frequently in the
calculation of various nuclear observables such as the mean
square radii, quadrupole moments, transition probabilities, \textit{etc}. Here we focus on lattice artifacts that produce some residual difference between the values in Eq.~(\ref{eq:matelonlat})
and Eq.~(\ref{eq:matelincon}) and the methods for removing them.

Let us start with the continuum limit. According to the Wigner-Eckart
theorem, the matrix element~(\ref{eq:matelincon}) can be transformed
into a product of two factors: Clebsch-Gordan (C-G) coefficients
and the
reduced matrix element that contains the essential non-trivial physics. Let $R_{1}(r)$ and $R_{2}(r)$ be the radial parts of the wave
functions $\phi_{l_{1}m_{1}}(\bm{r})$ and $\phi_{l_{2}m_{2}}(\bm{r})$, respectively,
so that
\begin{equation}
 \langle l_{1}m_{1}|r^{l^{\prime}}Y_{lm}|l_{2}m_{2}\rangle =  \langle l_{1} \vert r^{l^{\prime}} \vert l_{2} \rangle Q_{l_{2}m_{2},lm}^{l_{1}m_{1}} , \label{eq:Factorization}
\end{equation}
where
\begin{eqnarray}
\langle l_{1} \vert r^{l^{\prime}} \vert l_{2} \rangle &=& \int dr r^{l^{\prime}+2}R^*_{1}(r)R_{2}(r) \label{eq:radial} \\
Q_{l_{2}m_{2},lm}^{l_{1}m_{1}} & = & \int d\Omega Y_{l_{1}m_{1}}^{*}(\Omega)Y_{lm}(\Omega)Y_{l_{2}m_{2}}(\Omega).
\end{eqnarray}
The radial integral in Eq.~(\ref{eq:radial}) is the matrix element of the $l^{\prime}$-order
moment operator $ r^{l^{\prime}} $ and is independent of the
quantum numbers $m_{1}$, $m$
and $m_{2}$. Meanwhile $Q_{l_{2}m_{2},lm}^{l_{1}m_{1}}$ can be written as a product of C-G coefficients,
\begin{equation}
Q_{l_{2}m_{2},lm}^{l_{1}m_{1}} =  \sqrt{\frac{(2l+1)(2l_{2}+1)}{4\pi(2l_{1}+1)}}C_{l_{2}0,l0}^{l_{1}0}C_{l_{2}m_{2},lm}^{l_{1}m_{1}}.\label{eq:Qfactor}
\end{equation}
All of the dependence on the quantum numbers $m_{1}$, $m$,
$m_{2}$ and $l$ is absorbed into $Q_{l_{2}m_{2},lm}^{l_{1}m_{1}}$. Instead of simply the factoring out the C-G coefficient $C_{l_{2}m_{2},lm}^{l_{1}m_{1}}$, we find this factorization useful because the radial integrals of $ r^{l^{\prime}}$
are physically meaningful. In Table~\ref{tab:The-factor-}, we list
some factors $Q_{l_{2}m_{2},lm}^{l_{1}m_{1}}$ with $l_{1}=l_{2}=2$
and $l$ from $0$ to $4$. The other factors can be obtained
from standard tables of C-G coefficients. 

Since full rotational symmetry is broken on the lattice, it is not possible
to write the wave functions as products of radial and angular parts.
However, one can still apply the Wigner-Eckart theorem to each \textit{irrep}
 of the cubic group. As a result, the matrix elements calculated on the lattice belonging to the same \textit{ irreps } are related by C-G
coefficients of the cubic group, which can be computed easily using decompositions into spherical harmonics \cite{Altmann:1957,Rykhlinskaya2006_CPC174-903}.
For example, we have
\begin{equation}
	(2T_{2}1|r^{2}Y_{2E0}|2T_{2}1) = -\sqrt{\frac{1}{3}}(2T_{2}1|r^{2}Y_{2E2}|2T_{2}\bar{1}),
\end{equation}
where $Y_{2E0}$ and $Y_{2E2}$ are defined as
\begin{equation}
Y_{2E0}=Y_{20}, \; \; Y_{2E2}=\sqrt{\frac{1}{2}}(Y_{22}+Y_{2\bar{2}}), 
\end{equation}
in analogy to the relations in Eq.~(\ref{eq:Jeq2transformation}).  The ratio of $-\sqrt{1/3}$ is irrespective of the
lattice spacing, box size or the strength of the interaction. 

We will divide the lattice matrix elements in Eq.~(\ref{eq:matelonlat}) by the $Q_{l_{2}m_{2},lm}^{l_{1}m_{1}}$ as defined in Eq.~(\ref{eq:Qfactor}) whenever non-zero,
even though the factorization~in Eq.~(\ref{eq:Factorization}) is not exact
on the lattice. We represent the resulting quantity by double vertical
lines,
\begin{equation}
(l_{1}m_{1}\Vert r^{l^{\prime}}Y_{lm}\Vert l_{2}m_{2})=(l_{1}m_{1}|r^{l^{\prime}}Y_{lm}|l_{2}m_{2})/Q_{l_{2}m_{2},lm}^{l_{1}m_{1}},\label{eq:reducematrix}
\end{equation}
for $Q_{l_{2}m_{2},lm}^{l_{1}m_{1}} \ne 0$. It is clear that these reduced matrix elements all converge to the
radial matrix element $\langle l_{1} \vert r^{l^{\prime}} \vert l_{2} \rangle$ as $a\rightarrow0$.
At non-zero lattice spacing, however, the ratio will depend on
the  quantum numbers $m_1$, $m_2$ and $m$. Therefore, the splittings between different
components of Eq.~(\ref{eq:reducematrix}) are sensitive indicators
of the rotational symmetry breaking effects.

\begin{table}
\begin{centering}
\begin{tabular}{ccccccccccccccccc}
\hline 
$l_{1}$ & $l$ & $l_{2}$ &  &  &  & $m_{1}$ & $m$ & $m_{2}$ &  & $Q_{l_{2}m_{2},lm}^{l_{1}m_{1}}$ &  & $m_{1}$ & $m$ & $m_{2}$ &  & $Q_{l_{2}m_{2},lm}^{l_{1}m_{1}}$\tabularnewline
\hline 
\hline 
$2$ & $0$ & $2$ &  &  &  & $0$ & $0$ & $0$ &  & $\frac{1}{2\sqrt{\pi}}$ &  & $1$ & $0$ & $1$ &  & $\frac{1}{2\sqrt{\pi}}$\tabularnewline
 &  &  &  &  &  & $2$ & $0$ & $2$ &  & $\frac{1}{2\sqrt{\pi}}$ &  &  &  &  &  & \tabularnewline
 &  &  &  &  &  &  &  &  &  &  &  &  &  &  &  & \tabularnewline
$2$ & $2$ & $2$ &  &  &  & $0$ & $0$ & $0$ &  & $\frac{1}{7}\sqrt{\frac{5}{\pi}}$ &  & $2$ & $1$ & $1$ &  & $\sqrt{\frac{15}{14\pi}}$\tabularnewline
 &  &  &  &  &  & $2$ & $0$ & $2$ &  & $-\frac{1}{7}\sqrt{\frac{5}{\pi}}$ &  & $1$ & $1$ & $0$ &  & $\frac{1}{14}\sqrt{\frac{5}{\pi}}$\tabularnewline
 &  &  &  &  &  & $2$ & $2$ & $0$ &  & $-\frac{1}{7}\sqrt{\frac{5}{\pi}}$ &  & $1$ & $0$ & $1$ &  & $\frac{1}{14}\sqrt{\frac{5}{\pi}}$\tabularnewline
 &  &  &  &  &  & $1$ & $2$ & $\overline{1}$  &  & $-\sqrt{\frac{15}{14\pi}}$ &  &  &  &  &  & \tabularnewline
 &  &  &  &  &  &  &  &  &  &  &  &  &  &  &  & \tabularnewline
$2$ & $4$ & $2$ &  &  &  & $0$ & $0$ & $0$ &  & $\frac{3}{7\sqrt{\pi}}$ &  & $1$ & $3$ & $\overline{2}$ &  & $\frac{1}{2}\sqrt{\frac{5}{7\pi}}$\tabularnewline
 &  &  &  &  &  & $2$ & $0$ & $2$ &  & $\frac{1}{14\sqrt{\pi}}$ &  & $1$ & $2$ & $\overline{1}$ &  & $-\frac{1}{7}\sqrt{\frac{10}{\pi}}$\tabularnewline
 &  &  &  &  &  & $2$ & $1$ & $1$ &  & $-\frac{1}{14}\sqrt{\frac{5}{\pi}}$ &  & $1$ & $0$ & $1$ &  & $-\frac{2}{7\sqrt{\pi}}$\tabularnewline
 &  &  &  &  &  & $1$ & $1$ & $0$ &  & $\frac{1}{7}\sqrt{\frac{15}{2\pi}}$ &  & $2$ & $2$ & $0$ &  & $\frac{1}{14}\sqrt{\frac{15}{\pi}}$\tabularnewline
 &  &  &  &  &  & $2$ & $4$ & $\overline{2}$ &  & $\sqrt{\frac{5}{14\pi}}$ &  &  &  &  &  & \tabularnewline
\hline 
\end{tabular}
\par\end{centering}

\caption{\label{tab:The-factor-}The factor $Q_{l_{2}m_{2},lm}^{l_{1}m_{1}}$
defined in Eq.~(\ref{eq:Qfactor}) with $l_{1}=l_{2}=2$ and $l$
 from $0$ to $4$. The other ratios not listed here can be obtained
from standard tables of C-G coefficients.}
\end{table}

\subsection{\label{sub:2-d}Isotropic average}

We now focus on spatial anisotropies which are associated with the orientation of our lattice wave functions $\phi_{lm}$ relative to the lattice
axes. Let us illustrate this point by recalculating the matrix element
$(l_{1}m_{1}|r^{l^{\prime}}Y_{lm}|l_{2}m_{2})$ with a tilted lattice
which differs from the original one by a rigid body rotation. Let us assume
that the wave functions $\phi_{l_{1}m_{1}}$ and $\phi_{l_{2}m_{2}}$
can be smoothly interpolated in between lattice points as a function in continuous space.   Then we can define a tilted matrix element
as 
\begin{eqnarray}
 &  & (l_{1}m_{1}|r^{l^{\prime}}Y_{lm}|l_{2}m_{2})_{\Lambda}\nonumber \\
 & = & \sum_{\bm{n}}\phi_{l_{1}m_{1}}^{*}(R(\Lambda)\bm{n}a)|\bm{n}a|^{l^{\prime}}Y_{lm}(R(\Lambda)\hat{\bm{n}})\phi_{l_{2}m_{2}}(R(\Lambda)\bm{n}a),
\end{eqnarray}
where $\Lambda=(\alpha,\beta,\gamma)$ is a set of Euler angles and
$R(\Lambda)$ is an element of the SO(3) rotation group. In the continuum limit, rotational invariance guarantees that $(l_{1}m_{1}|r^{l^{\prime}}Y_{lm}|l_{2}m_{2})$
and $(l_{1}m_{1}|r^{l^{\prime}}Y_{lm}|l_{2}m_{2})_{\Lambda}$ are equal. For non-zero lattice spacing and non-zero angular
momenta, however, the tilted matrix element will depend on the quantity
$\Lambda$.  We would like to eliminate this unphysical orientation dependence from the final results.

A natural choice is to average the results over all possible
orientations, or equivalently, over the whole SO(3) group space. The
isotropically-averaged matrix element is defined as
\begin{eqnarray}
 &  & (l_{1}m_{1}|r^{l^{\prime}}Y_{lm}|l_{2}m_{2})_\circ=\int d^3\Lambda(l_{1}m_{1}|r^{l^{\prime}}Y_{lm}|l_{2}m_{2})_{{\rm \Lambda}}\nonumber \\
 & = & C_{l_{2}m_{2},lm}^{l_{1}m_{1}}\left[\frac{1}{2l_{1}+1}\sum_{m^{\prime},m_{1}^{\prime},m_{2}^{\prime}}C_{l_{2}m_{2}^{\prime},lm^{\prime}}^{l_{1}m_{1}^{\prime}}(l_{1}m'_{1}|r^{l^{\prime}}Y_{lm'}|l_{2}m'_{2})\right]. \label{eq:defineisoavg}
\label{eq:average} \end{eqnarray}
Here, $d^3\Lambda$ is the normalized invariant measure on the SO(3)
group space and the subscript ``$\circ$'' denotes the isotropic average.
In analogy with Eq.~(\ref{eq:reducematrix}), it is convenient to define a reduced isotropically-averaged matrix element,
\begin{equation}
(l_{1}\Vert r^{l^{\prime}}Y_{l}\Vert l_{2})_{\circ}=(l_{1}m_{1}|r^{l^{\prime}}Y_{lm}|l_{2}m_{2})_{\circ}/Q_{l_{2}m_{2},lm}^{l_{1}m_{1}}\label{eq:isotropicaverage}
\end{equation}
for $Q_{l_{2}m_{2},lm}^{l_{1}m_{1}} \ne 0$.  In the continuum limit $(l_{1}\Vert r^{l^{\prime}}Y_{l}\Vert l_{2})_{\circ}$ coincides with the radial matrix element $\langle l_{1} \vert r^{l^{\prime}} \vert l_{2} \rangle$.

We note that the radial matrix element $\langle l_{1} \vert r^{l^{\prime}} \vert l_{2} \rangle$ is not only independent of $m$, $m_1$ and $m_2$, but also independent of $l$. So a non-trivial test of  rotational symmetry  restoration is to check that $(l_{1}\Vert r^{l^{\prime}}Y_{l}\Vert l_{2})_{\circ}$ as defined in Eq.~(\ref{eq:isotropicaverage}) is independent of $l$. If $(l_{1}\Vert r^{l^{\prime}}Y_{l}\Vert l_{2})_{\circ}$ is to a good approximation independent of $l$, then we have succeeded in approximately factorizing the radial and angular parts of the lattice wave function by means of isotropic averaging. We will test this numerically
with the $\alpha$ cluster model in Sec.~\ref{sec:III}.

\section{\label{sec:III}Results}

\begin{figure}
\begin{centering}
\includegraphics[width=0.8\columnwidth]{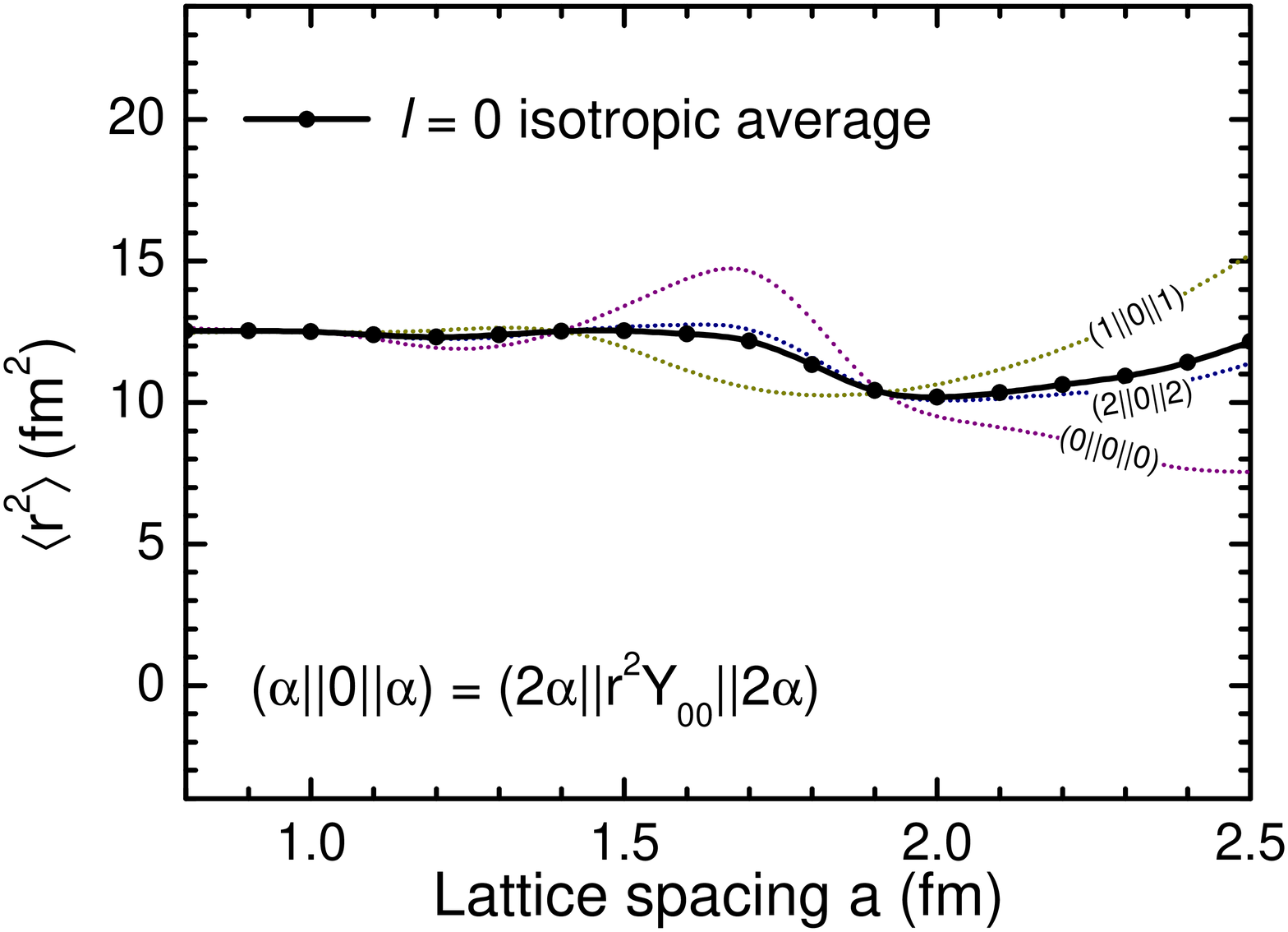}
\par\end{centering}

\begin{centering}
\includegraphics[width=0.8\columnwidth]{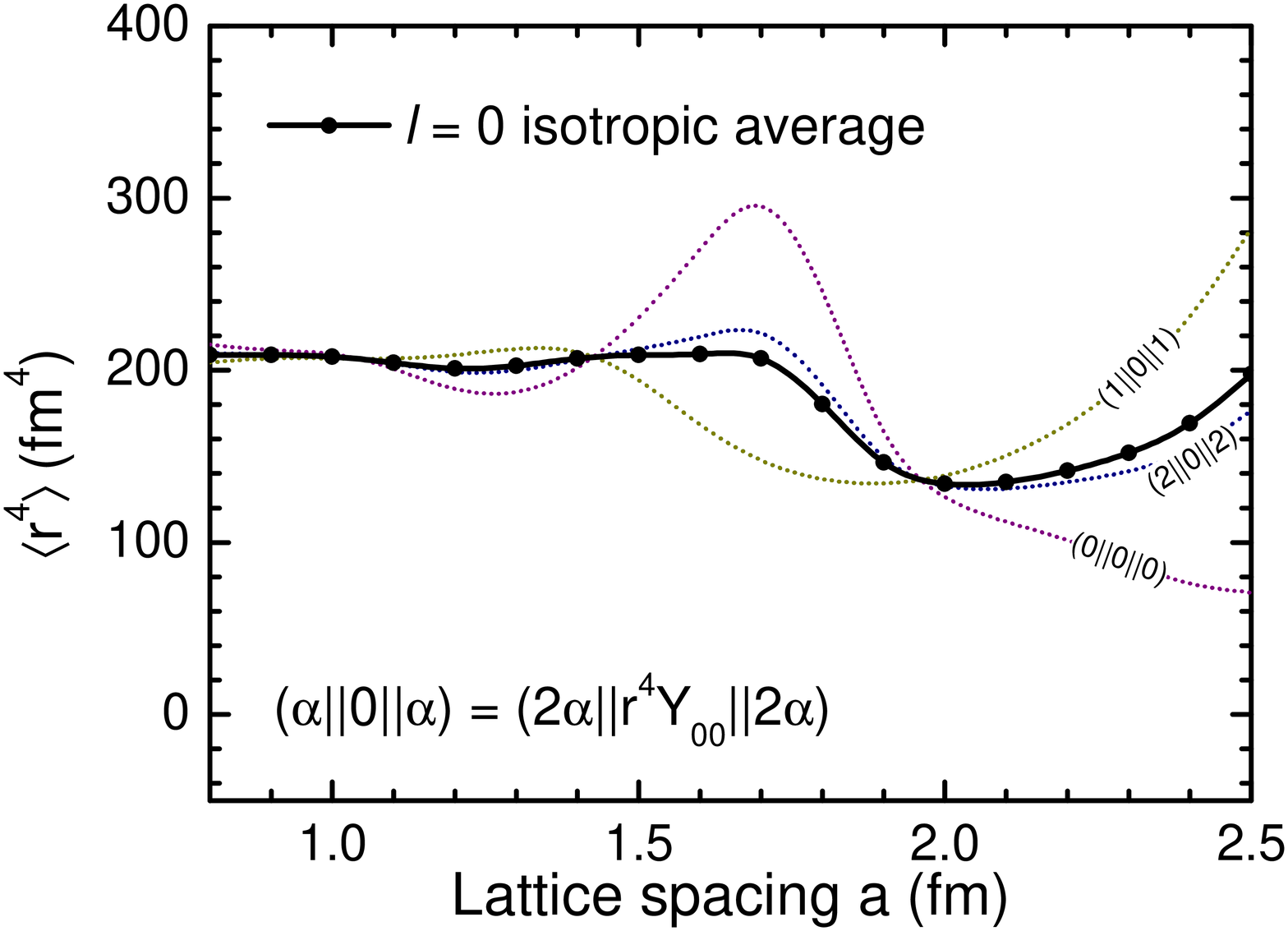}
\par\end{centering}

\caption{\label{fig:rsqandr4}(color online). \textit{Upper panel}: Mean square
radii $\langle r^{2}\rangle$ for the lowest $2^{+}$ states of the
$^{8}$Be nucleus as a function of $a$. $(\alpha\Vert0\Vert\alpha)$
is an abbreviation of the reduced lattice matrix element $(2\alpha\Vert r^{2}Y_{00}\Vert2\alpha)$
defined in Eq.~(\ref{eq:reducematrix}), which converges to $\langle
 r^{2} \rangle $
as $a\rightarrow0$. The box size $L$ is kept larger than 16 fm to
remove finite volume effects. The solid line represents the isotropic
average $(2\Vert r^{2}Y_{0}\Vert2)_\circ$ defined in Eq.~(\ref{eq:isotropicaverage}).
\textit{Lower panel}: Mean value $\langle  r^{4}  \rangle$ for the lowest
$2^{+}$ states of the $^{8}$Be nucleus as a function of $a$. $(\alpha\Vert0\Vert\alpha)$
is an abbreviation of the reduced lattice matrix element $(2\alpha\Vert r^{4}Y_{00}\Vert2\alpha)$,
which converges to $\langle r^{4}  \rangle$ as $a\rightarrow0$. The
solid line represents the isotropic average $(2\Vert r^{4}Y_{0}\Vert2)_\circ$.}
\end{figure}

We start with the mean square radius operator $r^{2}$. This corresponds
to setting $l=0$ and $l^{\prime}=2$ in Eq.~(\ref{eq:matelonlat})
and Eq.~(\ref{eq:matelincon}). In the upper panel of Fig.~\ref{fig:rsqandr4},
we show the expectation values of $r^{2}$ for the lowest $2^{+}$ states
as functions of the lattice spacing $a$. The eigenstate wave functions
$\psi_{2\tau k}$ are obtained by simultaneously diagonalizing the Hamiltonian $H$ in Eq.~(\ref{eq:continuumHamiltonian}) and the $R_{z}(\pi/2)$ operator.
We then construct the linear combinations $\phi_{2m}$ according to Eq.~(\ref{eq:Jeq2transformation}).
 We write $(m\Vert0\Vert m)$ as an abbreviation for $(2m\Vert r^{2}Y_{00}\Vert2m)$ and the solid curve denotes the
isotropic average $(2\Vert r^{2}Y_{00}\Vert2)_{\circ}$ defined in Eq.~(\ref{eq:isotropicaverage}).
Only three values with $m\geq0$ are shown since time reversal
symmetry ensures equal results for $m$ and $-m$. As discussed in Sec.~\ref{sec:II}, all these reduced matrix
elements converge to $\langle
 r^{2} \rangle \equiv\langle l_1=2\vert r^{2} \vert l_2=2 \rangle$ as $a\rightarrow0$.
Note that in all the following calculations, we remove the finite
volume effects by using a box size of $L\geq16$ fm.

The three branches are not linearly independent because of the cubic symmetries on the lattice. For scalar operators, the linear
relations among them are manifest. According to Eq.~(\ref{eq:Jeq2transformation}),
the wave functions $\phi_{21}$ and $\phi_{20}$ belong to the  \textit{irreps}
$E$ and $T_{2}$, respectively. Meanwhile the wave function
$\phi_{22}$ is an mixture of  the \textit{irreps} $E$ and $T_{2}$ with
equal weights. Thus, $(2\Vert0\Vert2)$ equals the arithmetic average of $(0\Vert0\Vert0)$ from \textit{irrep} $E$ and $(1\Vert0\Vert1)$ from \textit{irrep} $T_2$. So we find that the isotropically-averaged reduced matrix element $(2\Vert r^{2}Y_{0}\Vert2)_{\circ}$ is given by
\begin{equation}
(2\Vert r^{2}Y_{0}\Vert2)_{\circ}=\frac{3}{5}(1\Vert0\Vert1)+\frac{2}{5}(0\Vert0\Vert0),\label{eq:weightedaverage}
\end{equation}
where the numerators $3$ and $2$ are simply the dimensions
of each cubic representation. It is easy to verify that the weighted average formula is applicable
for any angular momentum with the factors $3$ and $2$ in Eq.~(\ref{eq:weightedaverage})
replaced by the corresponding \textit{irrep} dimensions. In Ref.~\cite{Lu2014_PRD90-034507}
we introduced the multiplet-weighted average to eliminate the anisotropic
effects in the bound state energies. We have now proven that this procedure
is equivalent to averaging  over lattice orientations and applies to
the expectation value of any scalar operator.

Next, let us examine the dependence of these reduced matrix elements
on the lattice spacing $a$. For $a\leq1.0$ fm, the three branches
merge, and for large $a$ they split and show oscillations. Before
discussing the details and the physics behind them, it is interesting
to compare Fig.~\ref{fig:rsqandr4} in this paper to Fig. 3 in Ref.~\cite{Lu2014_PRD90-034507}
where the calculated $2^{+}$ energies are shown as functions of $a$.
We immediately find that these figures look similar if we map $(1\Vert0\Vert1)$
to $E(2_{{\rm T}}^{+})$, and $(0\Vert0\Vert0)$ to $E(2_{{\rm E}}^{+})$.
More specifically, in Fig.~\ref{fig:rsqandr4} the splitting between
the two branches has two zeros near $a=1.4$ fm and $1.9$ fm. For
$a\leq1.0$ fm, the splitting is negligible. For $1.0\leq a\leq1.4$
fm, $(1\Vert0\Vert1)$ is higher than $(0\Vert0\Vert0)$. However,
in the region $1.4\leq a\leq1.9$ fm, the order is reversed. For $a\geq1.9$
fm, $(1\Vert0\Vert1)$ is once again higher and the splitting increases
monotonically. This behavior also occurs for the energies in Ref.~\cite{Lu2014_PRD90-034507}
with slightly different turning points. Additionally, the weighted
averages $(2\Vert r^{2}Y_{0}\Vert2)_\circ$ and $E(2_{{\rm A}}^{+})$
both show down bending in the transitional region $1.5\leq a\leq2.0$
fm, resulting in smaller expectation values for both energy and radii
at large lattice spacings.

The same pattern is also observed for other scalar operators. In the
lower panel of Fig.~\ref{fig:rsqandr4}, we show the results for
the $r^{4}$ operator which corresponds to setting $l=0$ and $l^{\prime}=4$.
Here, the symbols $(m\Vert0\Vert m)$ are abbreviations of $(2m\Vert r^{4}Y_{00}\Vert2m)$
and the solid curve denotes the isotropic average. All curves converge
to the expectation value $\langle r^{4}\rangle \equiv\langle l_1=2\vert r^{4} \vert l_2=2 \rangle$ as $a\rightarrow0$.
Again, the isotropic average is just the multiplet-weighted average
over the five-fold multiplet $\phi_{2m}$. Apparently, the oscillations
of these components as well as the down bending of the isotropic average
are quite similar to the ones observed for the $r^{2}$ operator. 

Based on the results shown in Fig.~\ref{fig:rsqandr4}, we can compare
the various components $(m\Vert0\Vert m)$ to the continuum limit
and select the ones with the least lattice spacing dependence. It seems that $(1\Vert0\Vert1)$
and $(0\Vert0\Vert0)$ are not optimal for both $r^{2}$ and $r^{4}$
operators because of their large deviations from the continuum values
for sparse lattices. The arithmetical average $(2\Vert0\Vert2)$ and
the multiplet-weighted average are close to each other for all lattice spacings
considered here. In principle, both of them can be used as a good
approximation to the continuum limit. Nevertheless, the multiplet-weighted
average is theoretically preferable because of its clear physical interpretation as isotropic
averaging.

Our conclusions for the operators $r^{2}$, $r^{4}$ as well as the
energy can be generalized straightforwardly to other scalar operators
on the lattice. For example, the linear relation among the components 
$(0\Vert0\Vert0)$, $(1\Vert0\Vert1)$ and  $(2\Vert0\Vert2)$ 
is also satisfied there. In order to estimate the continuum
value, we can calculate the isotropic average according to Eq.~(\ref{eq:isotropicaverage})
which equals the weighted average over the angular momentum multiplet
in the case of scalar operators. 

\begin{figure}
\begin{centering}
\includegraphics[width=0.8\columnwidth]{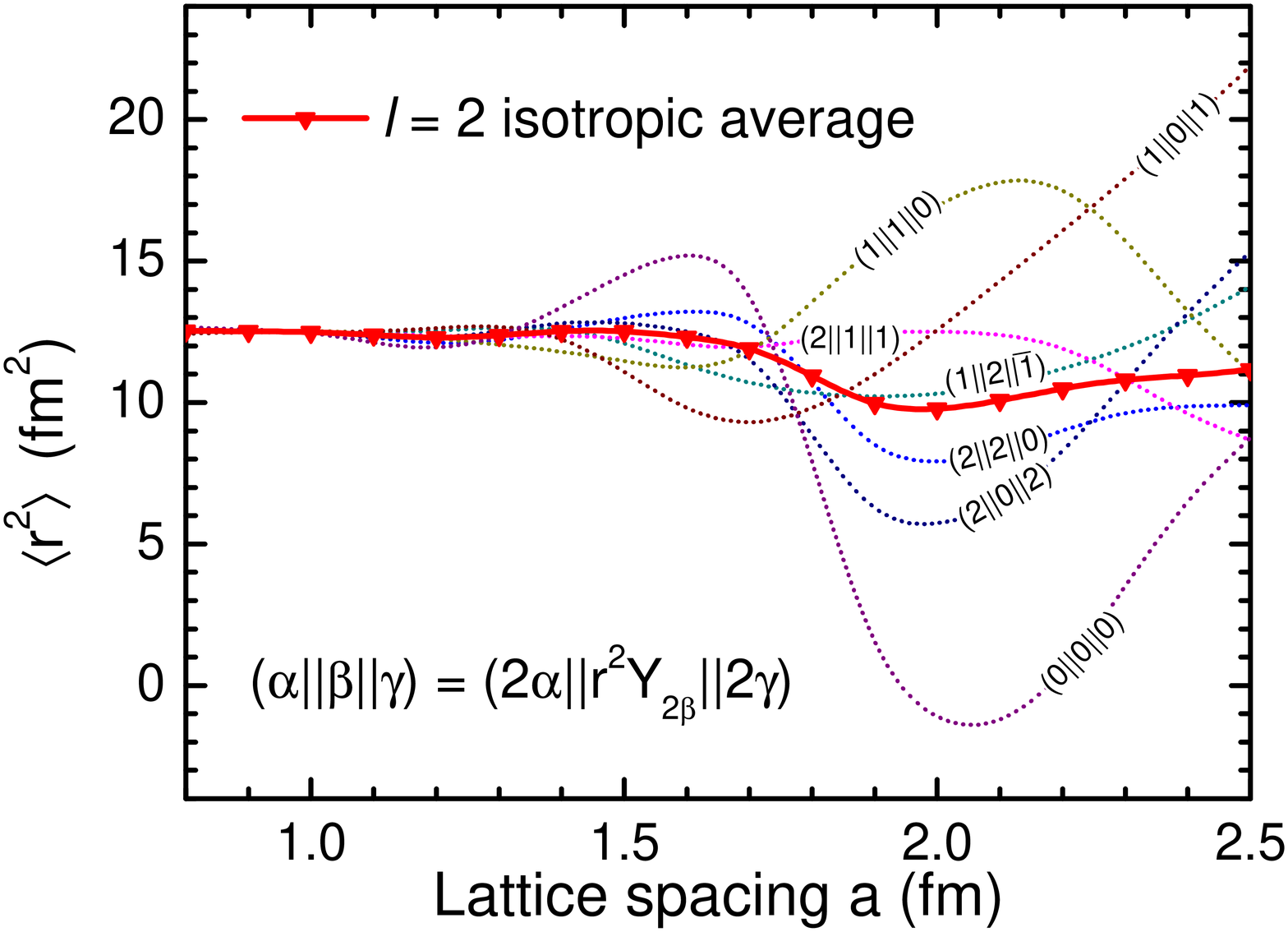}
\par\end{centering}

\begin{centering}
\includegraphics[width=0.8\columnwidth]{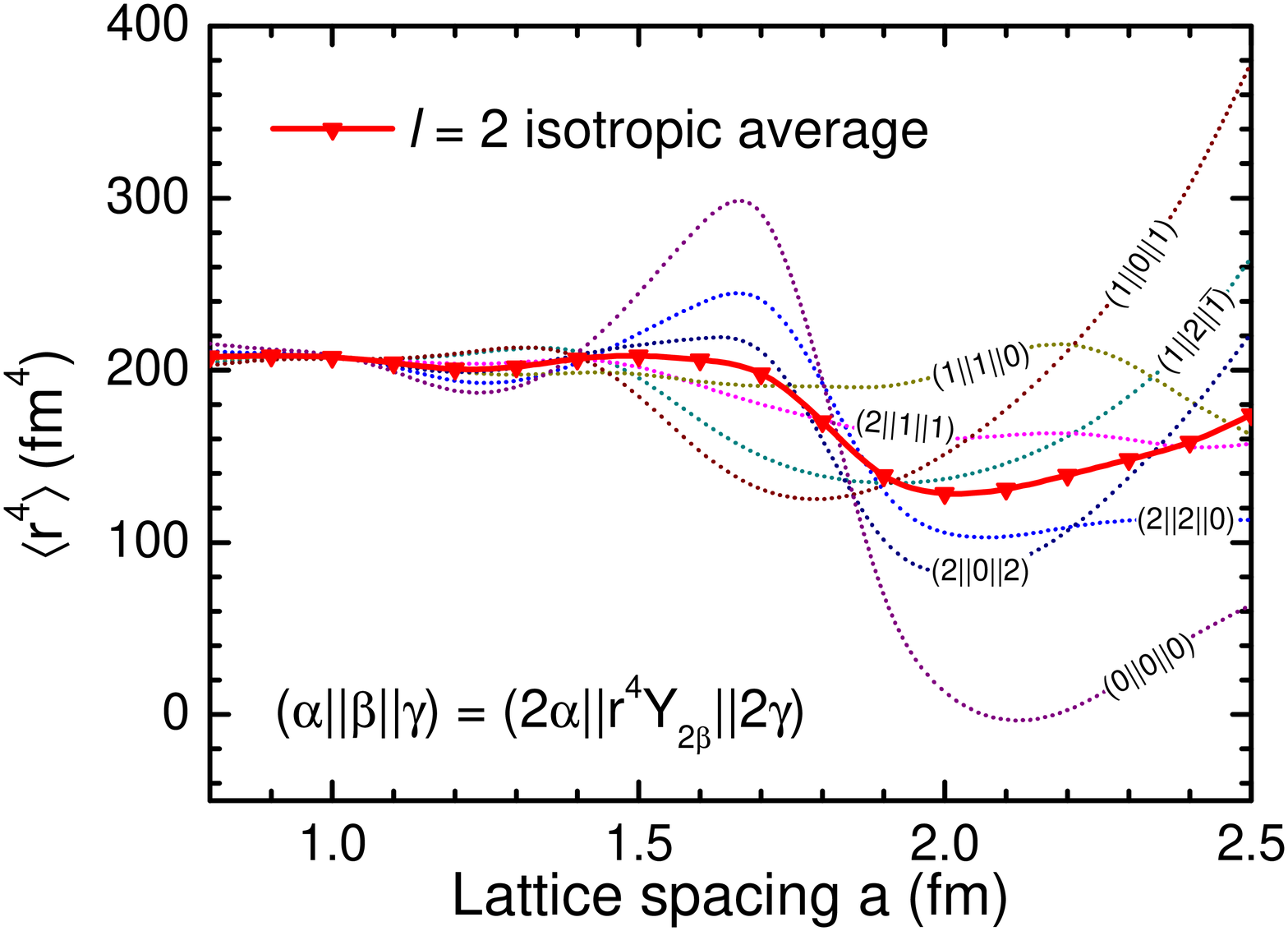}
\par\end{centering}

\caption{\label{fig:r2Y2andr4Y2}(color online). \textit{Upper panel}: Mean
square radii $\langle r^{2}\rangle$ for the lowest $2^{+}$ states
of the $^{8}$Be nucleus as a function of $a$. $(\alpha\Vert\beta\Vert\gamma)$
is an abbreviation of the reduced lattice matrix element $(2\alpha\Vert r^{2}Y_{2\beta}\Vert2\gamma)$
defined in Eq.~(\ref{eq:reducematrix}), which converges to $\langle r^{2}\rangle$
as $a\rightarrow0$. The solid line represents the isotropic average
$(2\Vert r^{2}Y_{2}\Vert2)_\circ$ defined in Eq.~(\ref{eq:isotropicaverage}).
\textit{Lower panel}: Mean value $\langle r^{4}\rangle$ for the lowest
$2^{+}$ states of the $^{8}$Be nucleus as a function of $a$. $(\alpha\Vert\beta\Vert\gamma)$
is an abbreviation of the reduced lattice matrix element $(2\alpha\Vert r^{4}Y_{2\beta}\Vert2\gamma)$,
which converges to $\langle r^{4}\rangle$ as $a\rightarrow0$. The
solid line represents the isotropic average $(2\Vert r^{4}Y_{2}\Vert2)_\circ$.}
\end{figure}

We now turn to the case of $l=2$. In the upper panel of Fig.~\ref{fig:r2Y2andr4Y2},
the symbols $(\alpha\Vert\beta\Vert\gamma)$ are abbreviations of
the reduced matrix elements $(2\alpha\Vert r^{2}Y_{2\beta}\Vert2\gamma)$
defined in Eq.~(\ref{eq:reducematrix}), where the subscripts $\alpha$,
$\beta$ and $\gamma$ run from $-2$ to $2$ and only the components
with $\alpha=\beta+\gamma$ are shown. The solid curve represents
the isotropic average defined in Eq.~(\ref{eq:isotropicaverage}).
All these curves converge to the mean square radius $\langle r^{2}\rangle$
as $a\rightarrow0$. Note that while the $2J+1$ wave functions in
an angular momentum multiplet are mixed with each other on the lattice,
some components with $\alpha\neq\beta+\gamma$ survive for large lattice
spacings. However, because the corresponding C-G coefficients vanish
in this case, these components do not contribute to the isotropic
average.

Compared to the case of the scalar operator $r^{2}$ shown in Fig.~\ref{fig:rsqandr4},
the insertion of the spherical harmonic $Y_{2\beta}$ makes Fig.~\ref{fig:r2Y2andr4Y2}
much more complicated. Nevertheless, we can still draw some general
conclusions. First, as for the scalar operators, we can show that
$(2\Vert0\Vert2)$ is just the arithmetical average of $(1\Vert0\Vert1)$
and $(0\Vert0\Vert0)$, while $(2\Vert2\Vert0)$ is just the arithmetical
average of $(1\Vert1\Vert0)$ and $(0\Vert0\Vert0)$. This point is
apparent in Fig.~\ref{fig:r2Y2andr4Y2}, if we notice that in each
group the three curves intersect at one point. Second, applying the
Wigner-Eckart theorem for the cubic group, we can write several linear
identities consisting of the lattice matrix elements. They concern not
only the components shown in Fig.~\ref{fig:r2Y2andr4Y2} but also
the one that vanish as $a\rightarrow0$.

In the upper panel of Fig.~\ref{fig:r2Y2andr4Y2}, most of the components show oscillations and have
more than one extremum in this region. For example, $(0\Vert0\Vert0)$
reaches a maximum at $a=1.6$ fm and two minima at $a=1.2$ fm and
$2.1$ fm, respectively, while the $(1\Vert0\Vert1)$ only has an
apparent minimum at $a=1.8$ fm. For large lattice spacings, the calculated
values usually deviate from the continuum ones by $50\%$ to $100\%$.
If we want to approximate the continuum limit using a single component
in the lattice calculations, $(2\Vert1\Vert1)$ is the best choice.
The corresponding curve only deviates slightly from the continuum
value 12.5 fm$^{2}$ when $a > 2.0$ fm, which is sufficient
for most recent lattice simulations of nuclei.

An anomaly occurs at $a=2.1$ fm where the calculated matrix element
$(0\Vert0\Vert0)$ becomes negative, whereas the continuum limit
is a definitely positive number $\langle r^{2}\rangle$. This  discrepancy arises because we do not calculate the expectation value of the operator $r^{2}$ with
the same wave functions on both sides, as was the case for scalar operators as shown in Fig.~\ref{fig:rsqandr4}.
For the $l=2$  matrix elements in Fig.~\ref{fig:r2Y2andr4Y2}, the reduced matrix element
$(0\Vert0\Vert0)$ is defined to be proportional to the expectation
value of the quadrupole operator $r^{2}Y_{20}$. On the lattice the angular part of the quadrupole
operator can not be separated completely, thus the insertion of the
spherical harmonic $Y_{20}$ is not fully canceled by the C-G coefficients
included in Eq.~(\ref{eq:reducematrix}). The remaining lattice
artifacts may become as large as the magnitude of the observable itself.
Therefore, it is dangerous to use randomly selected matrix elements
calculated at large $a$ as approximations of the continuum results,
otherwise we may even find vanishing results for certain lattice spacings,
which is unrealistic.

In spite of the diversity of the behaviors of the various components,
we can eliminate the anisotropy of the lattice artifacts by calculating
the isotropic average according to Eq.~(\ref{eq:isotropicaverage}).
In the upper panel of Fig.~\ref{fig:r2Y2andr4Y2}, the isotropic
average as a function of the lattice spacing is represented by the
solid line. In this case, the expression can not be written as a
simple multiplet-weighted average because the C-G coefficients are no longer all the same. Similar to the isotropic average shown in the upper panel of Fig.~\ref{fig:rsqandr4},
the isotropic average in Fig.~\ref{fig:r2Y2andr4Y2} curve slightly bends downward in the region $1.5\;{\rm fm}\leq a \leq 2.0\;
{\rm fm}$. For the lattice spacings considered here, the range of values
obtained are between 9.8~fm$^{2}$ and 12.5~fm$^{2}$, accounting
for a $20\%$ relative error with respect to the continuum limit.
The component that is closest to the isotropic average for all lattice
spacings is $(1\Vert2\Vert\bar{1})$.

Now let us change the radial factor of the inserted operator and keep
the angular part the same. In the lower panel of Fig.~\ref{fig:r2Y2andr4Y2},
the symbol $(\alpha\Vert\beta\Vert\gamma)$ denotes the lattice reduced
matrix element $(2\alpha\Vert r^{4}Y_{2\beta}\Vert2\gamma)$ which
converges to $\langle r^{4}\rangle$ as $a\rightarrow0$. The solid
line represents the isotropic average $(2\Vert r^{4}Y_{2}\Vert2)_\circ$.
Again the inserted spherical harmonic
$Y_{2\beta}$ is not fully canceled by the C-G coefficients on the
lattice. The remaining lattice artifacts shift the various components
from the continuum limit by different amounts for large lattice spacings.
Nevertheless, comparing the curves denoted by the same symbol in the
upper and lower panel of Fig.~\ref{fig:r2Y2andr4Y2}, we find that
their behavior is \textit{qualitatively} similar. For example, the
$(0\Vert0\Vert0)$ curves both show a maximum at $a=1.6$ fm and
a minimum at 2.1 fm, while the $(2\Vert0\Vert2)$ curves both
show a minimum at 2.0 fm. In other words, the magnitude of
the lattice artifacts may be different if the radial part of the inserted
operator is changed, but the pattern of oscillations and the sign
of the deviations are largely determined by the angular momenta of the
states and the irreducible tensor operator. 

\begin{figure}
\begin{centering}
\includegraphics[width=0.8\columnwidth]{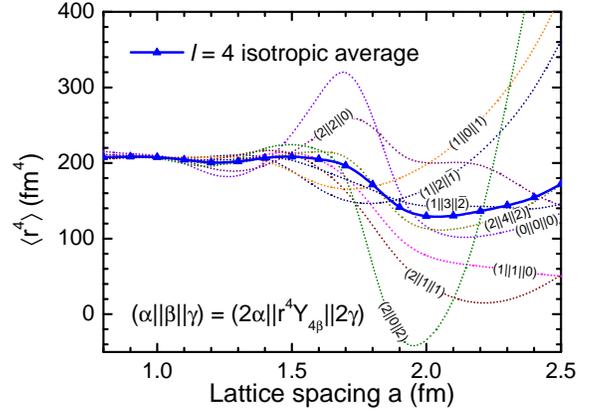}
\par\end{centering}

\caption{\label{fig:r4Y4}(color online) Mean value $\langle r^{4}\rangle$
for the lowest $2^{+}$ states of the $^{8}$Be nucleus as a function
of $a$. $(\alpha\Vert\beta\Vert\gamma)$ is an abbreviation of the
reduced lattice matrix element $(2\alpha\Vert r^{4}Y_{4\beta}\Vert2\gamma)$
defined in Eq.~(\ref{eq:reducematrix}), which converges to $\langle r^{4}\rangle$
as $a\rightarrow0$. The solid line represents the isotropic average
$(2\Vert r^{4}Y_{4}\Vert2)_\circ$ defined in Eq.~(\ref{eq:isotropicaverage}). }
\end{figure}

Next, we turn to the irreducible tensor operators with $l=4$. In
Fig.~\ref{fig:r4Y4} the symbol $(\alpha\Vert\beta\Vert\gamma)$
denotes the lattice reduce matrix element $(2\alpha\Vert r^{4}Y_{4\beta}\Vert2\gamma)$
which converges to $\langle r^{4}\rangle$ as $a\rightarrow0$. The
solid line represents the isotropic average $(2\Vert r^{4}Y_{4}\Vert2)_\circ$.
Now the number of independent components is larger than that in the
case of $l=2$ and the situation is different. For example, the $(0\Vert0\Vert0)$
curve is much closer to the isotropic average compared to the corresponding
curve in the lower panel of Fig.~\ref{fig:r2Y2andr4Y2}, while the
$(2\Vert0\Vert2)$ curve has a pronounced minimum at $a=2.0$
fm. The component closest to the continuum limit as well as the isotropic
average for all lattice spacings considered here is $(1\Vert3\Vert\bar{2})$.
There also exist a number of linear relations among these components,
which is a consequence of the remaining unbroken symmetry group. There
is also a negative component $(2\Vert0\Vert2)$ at $a=2.0$ fm.
We will not discuss all the details because most of them are natural
extensions of the $l=2$ case.

\begin{figure}
\begin{centering}
\includegraphics[width=0.8\columnwidth]{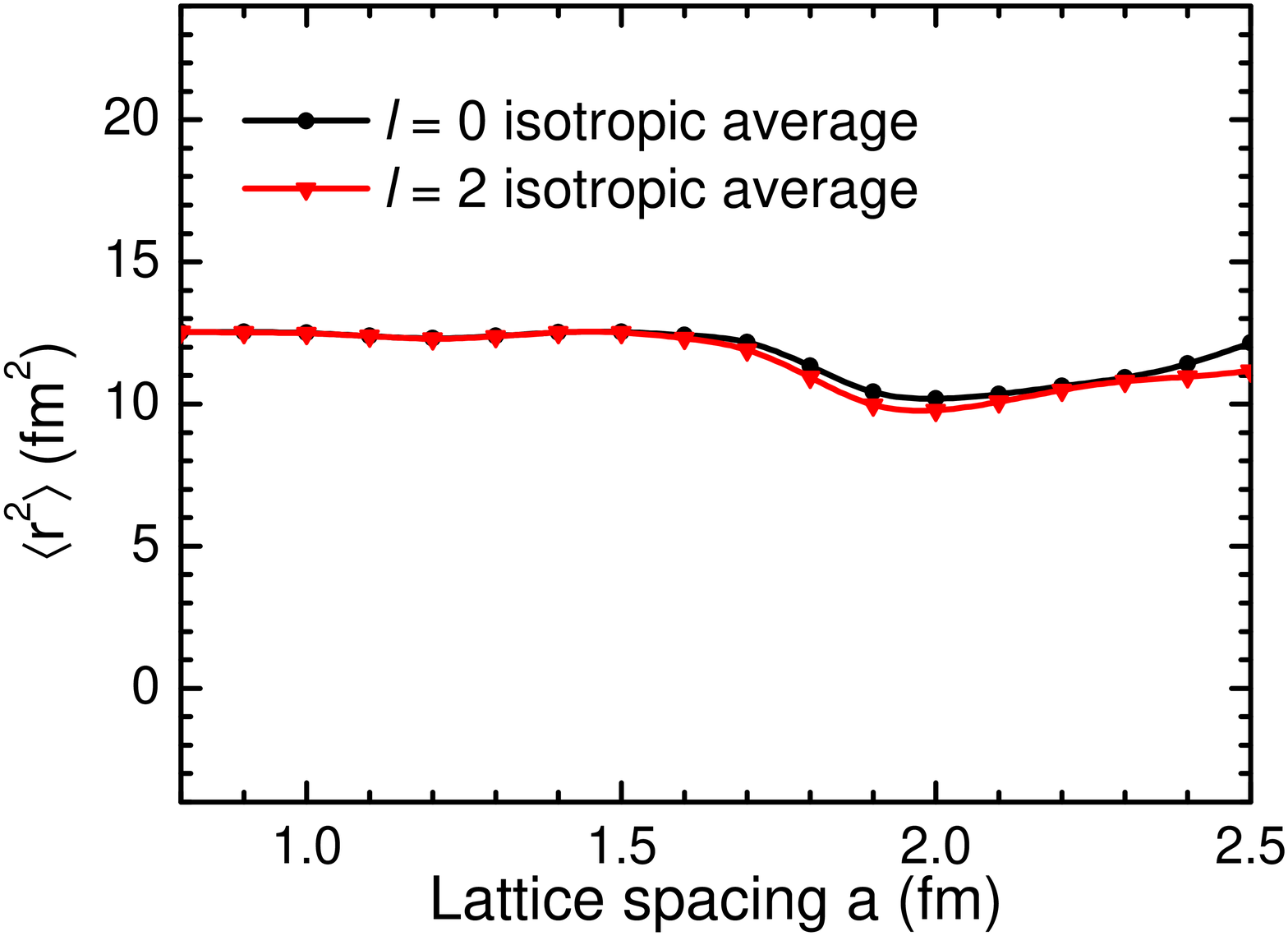}
\par\end{centering}

\begin{centering}
\includegraphics[width=0.8\columnwidth]{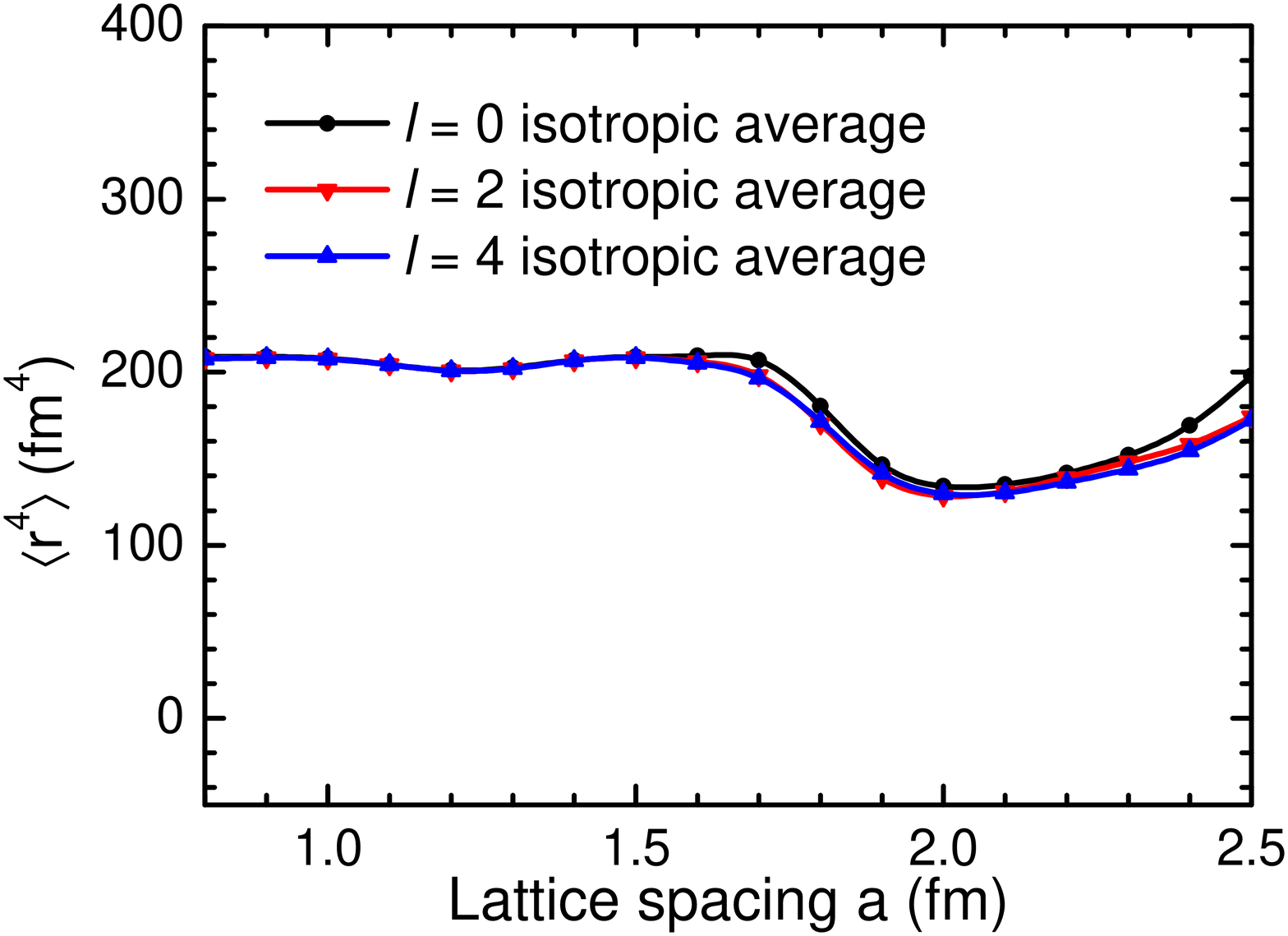}
\par\end{centering}

\caption{\label{fig:compare}(Color online) \textit{Upper panel}: Mean square
radii $\langle r^{2}\rangle$ for the lowest $2^{+}$ states of the
$^{8}$Be nucleus as a function of $a$. The black and red lines represent
the isotropic average $(2\Vert r^{2}Y_{0}\Vert2)_\circ$ and $(2\Vert r^{2}Y_{2}\Vert2)_\circ$,
respectively. \textit{Lower panel}: Expectation value $\langle r^{4}\rangle$
for the lowest $2^{+}$ states of the $^{8}$Be nucleus as a function
of $a$. The black, red and blue lines represent the isotropic averages
$(2\Vert r^{4}Y_{0}\Vert2)_\circ$, $(2\Vert r^{4}Y_{2}\Vert2)_\circ$
and $(2\Vert r^{4}Y_{4}\Vert2)_\circ$, respectively. }
\end{figure}

Finally, let us examine the postulate that the anisotropy resulting
from the lattice artifacts is removed in the isotropic average, Eq.~(\ref{eq:isotropicaverage}).
In the upper panel of Fig.~\ref{fig:compare} we show the comparison
between the isotropic averages $(2\Vert r^{2}Y_{0}\Vert2)_\circ$ and
$(2\Vert r^{2}Y_{2}\Vert2)_\circ$. The former one is calculated with
a simple multiplet averaging over the five fold branches, while the
latter one is obtained by a relatively complicated averaging with
the C-G coefficients as the weights. Clearly, for all the lattice
spacings considered here, the difference between the two curves is
rather small. Especially, the down bending occurs for the same lattice
spacing and the magnitudes are also similar. Remembering that the
C-G coefficients are included explicitly in the definition of isotropic
average Eq.~(\ref{eq:isotropicaverage}), we conclude that the effect
of the angular part of the inserted operators is canceled by the C-G
coefficients in the nominator. As is discussed in the previous section,
this is a strong evidence for approximate rotational symmetry restoration and that we have factorized the radial and angular parts of the lattice wave function by means of isotropic averaging.

In the lower panel of Fig.~\ref{fig:compare} we show the quantities
that converge to $\langle r^{4}\rangle$ as $a\rightarrow0$, including
$(2\Vert r^{4}Y_{0}\Vert2)_\circ$, $(2\Vert r^{4}Y_{2}\Vert2)_\circ$
and $(2\Vert r^{4}Y_{4}\Vert2)_\circ$. The three curves coincide with
each other even for $a > 2.0$ fm, which means
that the rotational symmetry is restored to a large extent after the
isotropic averaging. In particular, the difference between the $l=2$
and $l=4$ results is almost zero for all the lattice spacings.

\section{summary}

In lattice calculations, discretization errors can break  rotational
symmetry. The degeneracy of bound state multiplets with the same angular
momentum is broken on the lattice. The resulting wave functions are
classified according to the irreducible representations\textit{ }of
the cubic group instead of the full SO(3) rotational group. For most of the observables represented by the irreducible tensor
operators, the relations between the various components become complicated.
In this paper, we used an $\alpha$ cluster model to investigate the
lattice matrix elements of such operators. We have shown that the
qualitative behaviors of the various matrix elements versus lattice spacing
is mainly determined by the angular momenta of the states and operators. The matrix elements of different operators with the same angular momentum
show a similar behavior as functions of the lattice spacing.

We have also defined an isotropic average in which the various components
of the matrix element are linearly combined. The weight for a component
is proportional to the corresponding C-G coefficients with the same
quantum numbers. We have shown that such a calculation is equivalent
to averaging over all possible lattice orientations.
In such a calculation, we eliminate to a good approximation the anisotropy of the lattice artifacts. This point is illustrated by numerical calculations
for the $^{8}$Be nucleus with the $\alpha$ cluster model. 

We have
calculated the isotropic averages as functions of the lattice spacing
for irreducible tensor operators with different angular momenta and
same radial factors and find good agreement with continuum limit values. Comparing the results calculated with different lattice spacings,
we found that there is a down bending of $\langle r^2 \rangle$ and  $\langle r^4 \rangle$ in the region 1.7
fm $\leq a\leq$ 2.0 fm. For $a < 1.7$ fm, the lattice artifacts are very small.

Although all the conclusions in this paper are obtained with a simple
$\alpha$ cluster model, the results can be applied immediately to
\textit{ab initio} lattice simulations as well. For example, when calculating the transition amplitude between
low-energy excited states and the ground state of a nucleus, we can
improve the results by calculating the isotropic average to eliminate rotational symmetry breaking effects on the lattice.

\begin{acknowledgments}
We acknowledge partial financial support from the Deutsche Forschungsgemeinschaft
(Sino-German CRC 110), the Helmoholtz Association (Contract No. VH-VI-417),
BMBF (Grant No. 05P12PDTEE), the U.S. Department of Energy (DE-FG02-03ER41260),
by the EU HadronPhysics3 project and the Magnus Ehrnrooth Foundation
of the Finnish Society of Sciences and Letters.
\end{acknowledgments}
\bibliographystyle{apsrev4-1}
%

\end{document}